\documentclass[twoside,twocolumn,9pt]{article}
\usepackage{extsizes}
\usepackage[super,sort&compress,comma]{natbib} 
\usepackage[version=3]{mhchem}
\usepackage[left=1.5cm, right=1.5cm, top=1.785cm, bottom=2.0cm]{geometry}
\usepackage{balance}
\usepackage{mathptmx}
\usepackage{sectsty}
\usepackage{graphicx} 
\usepackage{lastpage}

\usepackage[format=plain,justification=justified,singlelinecheck=false,font={stretch=1.125,small,sf},labelfont=bf,labelsep=space]{caption}
\usepackage{float}
\usepackage{fancyhdr}
\usepackage{fnpos}
\usepackage[english]{babel}
\addto{\captionsenglish}{%
  
}
\usepackage{array}
\usepackage{droidsans}
\usepackage{charter}
\usepackage[T1]{fontenc}
\usepackage[usenames,dvipsnames]{xcolor}
\usepackage{setspace}
\usepackage[compact]{titlesec}
\usepackage{hyperref}

\hypersetup{
    colorlinks = true, 
    linkcolor = blue, 
    citecolor = blue
}
\usepackage{booktabs}
\usepackage{multirow}

\usepackage{amsmath}
\newcommand{\abs}[1]{\left\lvert#1\right\rvert}
\newcommand{\norm}[1]{\left\lVert#1\right\rVert}

\usepackage{amsthm}
\newtheorem*{theorem*}{Theorem}
\theoremstyle{definition}
\newtheorem{proofpart}{Part}
\makeatletter
\@addtoreset{proofpart}{theorem*}
\makeatother

\definecolor{cream}{RGB}{222,217,201}

\begin{document}

\pagestyle{fancy}
\thispagestyle{plain}
\fancypagestyle{plain}{
\renewcommand{\headrulewidth}{0pt}
}

\makeFNbottom
\makeatletter
\renewcommand\LARGE{\@setfontsize\LARGE{15pt}{17}}
\renewcommand\Large{\@setfontsize\Large{12pt}{14}}
\renewcommand\large{\@setfontsize\large{10pt}{12}}
\renewcommand\footnotesize{\@setfontsize\footnotesize{7pt}{10}}
\makeatother

\renewcommand{\thefootnote}{\fnsymbol{footnote}}
\renewcommand\footnoterule{\vspace*{1pt}%
\color{cream}\hrule width 3.5in height 0.4pt \color{black}\vspace*{5pt}} 
\setcounter{secnumdepth}{5}

\makeatletter 
\renewcommand\@biblabel[1]{#1}            
\renewcommand\@makefntext[1]%
{\noindent\makebox[0pt][r]{\@thefnmark\,}#1}
\makeatother 
\renewcommand{\figurename}{\small{Fig.}~}
\sectionfont{\sffamily\Large}
\subsectionfont{\normalsize}
\subsubsectionfont{\bf}
\setstretch{1.125} 
\setlength{\skip\footins}{0.8cm}
\setlength{\footnotesep}{0.25cm}
\setlength{\jot}{10pt}
\titlespacing*{\section}{0pt}{4pt}{4pt}
\titlespacing*{\subsection}{0pt}{15pt}{1pt}

\fancyfoot{}
\fancyfoot[LO,RE]{\vspace{-7.1pt}\includegraphics[height=9pt]{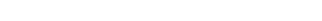}}
\fancyfoot[CO]{\vspace{-7.1pt}\hspace{13.2cm}\includegraphics{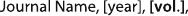}}
\fancyfoot[CE]{\vspace{-7.2pt}\hspace{-14.2cm}\includegraphics{head_foot/RF}}
\fancyfoot[RO]{\footnotesize{\sffamily{1--\pageref{LastPage} ~\textbar  \hspace{2pt}\thepage}}}
\fancyfoot[LE]{\footnotesize{\sffamily{\thepage~\textbar\hspace{3.45cm} 1--\pageref{LastPage}}}}
\fancyhead{}
\renewcommand{\headrulewidth}{0pt} 
\renewcommand{\footrulewidth}{0pt}
\setlength{\arrayrulewidth}{1pt}
\setlength{\columnsep}{6.5mm}
\setlength\bibsep{1pt}

\makeatletter 
\newlength{\figrulesep} 
\setlength{\figrulesep}{0.5\textfloatsep} 

\newcommand{\topfigrule}{\vspace*{-1pt}%
\noindent{\color{cream}\rule[-\figrulesep]{\columnwidth}{1.5pt}} }

\newcommand{\botfigrule}{\vspace*{-2pt}%
\noindent{\color{cream}\rule[\figrulesep]{\columnwidth}{1.5pt}} }

\newcommand{\dblfigrule}{\vspace*{-1pt}%
\noindent{\color{cream}\rule[-\figrulesep]{\textwidth}{1.5pt}} }

\makeatother

\twocolumn[
  \begin{@twocolumnfalse}
{\includegraphics[height=30pt]{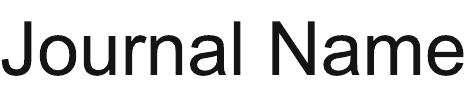}\hfill\raisebox{0pt}[0pt][0pt]{\includegraphics[height=55pt]{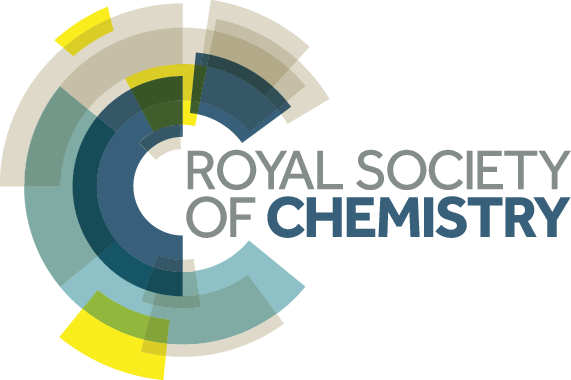}}\\[1ex]
\includegraphics[width=18.5cm]{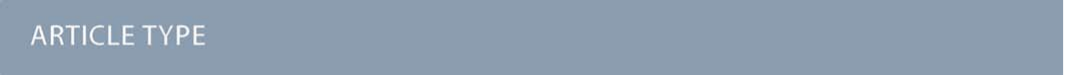}}\par
\vspace{1em}
\sffamily
\begin{tabular}{m{4.5cm} p{13.5cm} }

\includegraphics{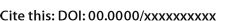} & \noindent\LARGE{\textbf{Towards an information-based theory of structure}} \\
\vspace{0.3cm} & \vspace{0.3cm} \\
 & \noindent\large{Glenn D. Hibbard and John \c{C}amk{\i}ran} \\

\includegraphics{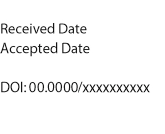} & \noindent\normalsize{We call for a theory of the particle-scale structure of materials that is based on the general notion of information rather than its special case of symmetry. An inherent limitation to the symmetry-based understanding of structure is described. The rapid decay in interaction strength with interparticle distance is used to argue for the representability of a system locally through the neighbourhoods of its constituent particles. The extracopularity coefficient $E$ is presented as a local quantifier of information and compared to point group order $\abs{G}$, a local quantifier of symmetry. The former is found to have nearly double the resolution of the latter for a set of commonly encountered coordination geometries. A proof is given for the generality of extracopularity over point symmetry. Some practical challenges and future perspectives are discussed.} \\

\end{tabular}

 \end{@twocolumnfalse} \vspace{0.6cm}

  ]

\renewcommand*\rmdefault{bch}\normalfont\upshape
\rmfamily
\section*{}
\vspace{-1cm}


\footnotetext{\textit{Department of Materials Science and Engineering, University of Toronto, 184 College St, Toronto, ON M5S 3E4, Canada. E-mail: john.camkiran@utoronto.ca}}





\section*{Symmetry versus information}
Much of the prevailing theory of the structure of materials at the particle scale hinges upon the notion of symmetry.\cite{degraef_2012} Indeed, the discrete spatial symmetries characteristic of crystalline matter give rise to an elegant structural theory that has seen great success in predicting a long list of physical phenomena.\cite{nye_1985,bowman_2004} A widely acknowledged shortcoming of this symmetry-based understanding, however, is that it offers limited insight into the structure of non-crystalline media\cite{bernal_1959, tanaka_2019} -- an area of growing scientific interest \cite{berthier_2023} and technological importance.\cite{gao_2022} The origin of this shortcoming lies in the fact that symmetry merely constitutes a special case of a more general mathematical notion that encompasses all forms of regularity inherent to a data source, often called \mbox{\textit{information}}.\cite{cover_2012} In this article, we consider the possibility of a general theory of materials structure in which the notion of information plays the role that has traditionally been played by symmetry.

\section*{From positions to distances}

Density functional theory reveals that the macroscopic properties of a solid are uniquely determined by its electronic density in ground state and that this density is obtainable from the external potential due to positively charged atomic nuclei.\cite{cohen_2016} Supposing the charges of the nuclei are known, the potential is in turn recoverable from nuclear positions via Coulomb’s law. Thus, the ground-state properties of a solid are fully determined by the positions of its nuclei. This provides a physical rationale for the mathematical representation of a material as a set of point-like particles per crystallography and the classical formulations of statistical mechanics and molecular dynamics.

But while particle positions are what formally specify the state of a system, it is the distances between those particles that constitute the fundamental quantity underlying structure. This follows from the fact that particle positions can be recovered from interparticle distances up to translation, rotation, and reflection,\cite{cox_1994} which have no effect on the total energy of a system. Indeed, the information content of particle positions is simply the information content of interparticle distances plus some extraneous information regarding the system’s overall position, orientation, and handedness. It is possible to express this relationship symbolically using Shannon's information entropy functional $H$ as \mbox{follows}:\cite{cover_2012}
\begin{equation}
H(P) = H(D) + H(T), 
\label{eq:position-distance}
\end{equation}
where $P$ denotes positions, $D$ distances, and $T$ an isometric transformation (i.e. a translation, a rotation, and a reflection).

\section*{From distances to bond angles}

A property common to all interactions among particles of atomic size of greater is that their strength decays rapidly with interparticle distance. As a direct result of this property, the structure of a particle system manifests most strongly in the neighbourhoods of its constituent particles.
 
\begin{figure}[!t]
    \centering
    \includegraphics[width=\linewidth]{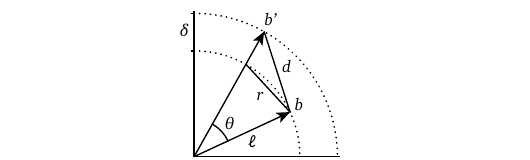}
    \caption{As the bond length difference $\delta$ approaches $0$, the interparticle distance $d$ tends to the chord length $r$. But $r$ depends on the bond angle $\theta$ through the relation ${r = 2 \ell \sin(\theta/2)}$ up to a scaling by $\ell$. Thus, the distance $d$ is determined approximately by the angle $\theta$.}
    \label{fig:trigonometry}
\end{figure}

Particle neighbourhoods tend to be approximately spherical (i.e. characterised by neighbours nearly equidistant from the central particle) as a consequence of the isotropy of space. By virtue of this approximate sphericality, the distance between any two neighbours of a given particle 
depends almost entirely on the angle between the vectors that indicate their position relative to the central particle; such vectors are commonly called \textit{bonds},\cite{steinhardt_1983} and the angles between them, \textit{bond angles}.\cite{ackland_2006} \mbox{Fig. \ref{fig:trigonometry}} illustrates this dependence geometrically.

\newpage
Just as with the position--distance relationship, the distance--angle relationship can be captured using the functional $H$. Observe that the triple consisting of bond angles $\Theta$, bond length differences $\Delta$, and the nearest-neighbour distance $\ell$ fully determines and is determined by interparticle distances $D$. These quantities are thereby equivalent in information,
\begin{align}
    H(D) = H(\Theta, \Delta, \ell).
\end{align}
Twice applying the chain rule\cite{cover_2012} to the right-hand side, then invoking the independence between $\Theta$ and $\ell$, we get the following general relationship:
\begin{equation}
    H(D) = H(\Theta) + H(\Delta \mid \ell, \Theta) + H(\ell).
\end{equation}

At the spherical neighbourhood limit, vanishing bond length differences render $H(\Delta) = 0$. By monotonicity, $H(\Delta \mid \ell, \Theta) \leq H(\Delta)$, and nonnegativity, $H(\Delta \mid \ell, \Theta) \geq 0$, we have $H(\Delta \mid \ell, \Theta) = 0$.\cite{cover_2012} Thus, interparticle distances $D$ are informationally equivalent to bond angles $\Theta$, up to a contribution from the nearest-neighbour distance $\ell$, 
\begin{equation}
    H(D) = H(\Theta) + H(\ell).
\end{equation}
In systems with a simple crystalline ground state, $H(\ell)$ tends to $0$ with increasing density, leaving $H(D) = H(\Theta)$. This underscores the importance of bond angles to structure at the particle scale.

\section*{Extracopularity and $E$}

From elementary combinatorics, we know that a particle with $k$ bonds has at most $(k^2-k)/2$ different bond angles. A key observation regarding energetically favourable neighbourhood geometries is that they exhibit a much lower diversity in bond angles than combinatorics would allow. Such geometries tend to resemble convex polyhedra with regular-polygonal faces,\cite{malins_2013_b} a few of which are illustrated in \mbox{Fig. \ref{fig:12-coordinate-geometries}}. 

Consider, for instance, the icosahedral geometry, which is of special importance to systems like supercooled liquids\cite{chang_2023} and metallic glasses.\cite{kumar_2019} Observe that the $66$ bond pairs implied by its $12$ bonds make only three different angles, namely  $\sim 63.4^\circ$, $\sim 116.6^\circ$, and $180^\circ$. These pairs are redundant in the sense that only three different angles suffice to describe all $66$ of them; they are thus losslessly compressible.\cite{cover_2012} Repeating this exercise for the three other geometries in Fig. \ref{fig:12-coordinate-geometries}, the number of different bond angles is found to be a quantity able to distinguish between the various ways of arranging $12$ neighbours around a particle.

\newpage
In a previous work,\cite{camkiran_2022a} the occurrence of fewer different bond angles than combinatorially possible was termed \textit{extracopularity}. This phenomenon was quantified by a coefficient $E$, defined as the conditional Hartley entropy \cite{hartley_1928} of bond pairs given bond angles,
\begin{equation}
    E = \log_2(n) - \log_2(m), \quad n>0,
\end{equation}
where $n$ is the number of bond pairs and $m$ is the number of different bond angles. The extracopularity coefficient is often given more explicitly in terms of the coordination number $k$ by
\begin{equation}
        E = \log_2 \left( \frac{k^2-k}{2m} \right), \quad k>1.
\end{equation}
As a tool, $E$ has found early application in the studies of convex polyhedra,\cite{camkiran_2023} cellular materials,\cite{choukir_2023} and radiation damage.\cite{stimac_2023} 

\section*{Local structure via symmetry and information}

\begin{figure*}
 \centering
 \includegraphics[width=\textwidth]{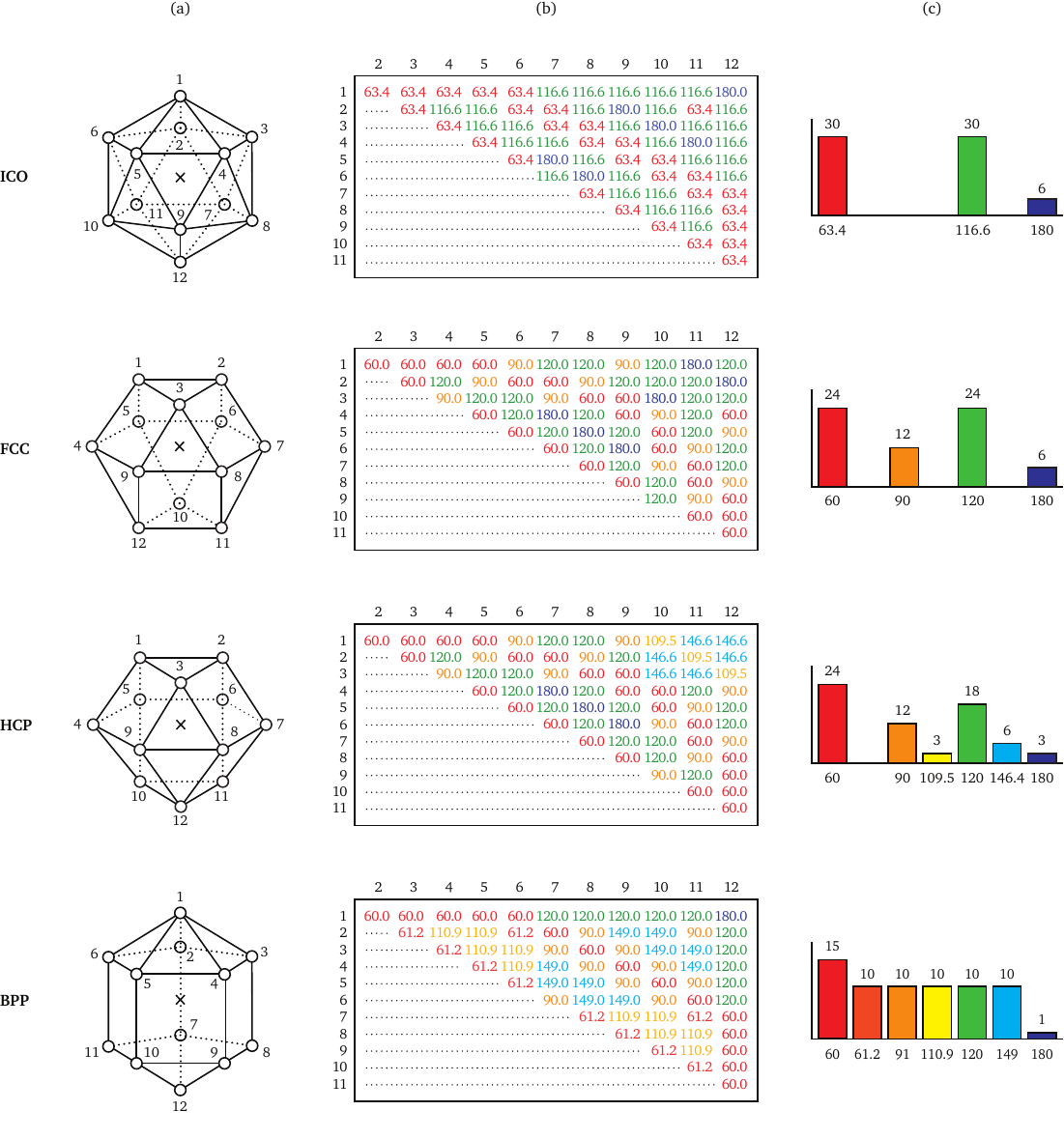}
 \caption{A comparison of the icosahedral (ICO), cuboctahedral (FCC), anticuboctahedral (HCP), and bicapped pentagonal prismatic (BPP) geometries by (a) polyhedral diagram, (b) bond-angle table, and (c) bond-angle frequency plot. Notably, FCC and HCP are related by a $60$-degree rotation of the lower cupola about the central vertical axis, while ICO and BPP are related by a $36$-degree rotation in the lower pentagonal pyramid about that same axis. In (b), lower triangular elements are omitted on account of their redundancy with the upper triangular elements.}
 \label{fig:12-coordinate-geometries}
\end{figure*}

The phenomenon of extracopularity, as quantified by $E$, embodies the information-based approach to local structure in the same way that the phenomenon of point symmetry, as quantified by point group order $\abs{G}$, embodies the symmetry-based approach. A basic performance criterion for local approaches to structural analysis is the extent to which distinct coordination geometries can be distinguished. Table \ref{table:12-coordinate-geometries} presents $E$ and $\abs{G}$ for each of the four $12$-coordinate geometries studied in \mbox{Fig. \ref{fig:12-coordinate-geometries}}. Observe that the ranking of these geometries by $E$ resembles their ranking by $\abs{G}$. This suggests that the symmetry- and information-based approaches to local structural analysis are comparable in analytical power for $12$-coordinate geometries.

Where the advantage of the information-based approach shows is in the analysis of coordination geometries that are symmetrically identical. Table \ref{table:cubic-geometries} presents $E$ and $\abs{G}$ for cubic geometries, which are characterised by the point group $O_h$.  Here, $\abs{G}$ fails at the most basic level, being unable to make any distinction between those coordination geometries much less rank them on an ordinal scale.\cite{stevens_1946} By contrast, $E$ succeeds in both respects. The ranking implied by $E$ is further observed to agree with that of the familiar atomic packing fraction ($\eta_{\, \text{FCC}} = 0.74$, $\eta_{\, \text{BCC}} = 0.68$, $\eta_{\, \text{SC}} = 0.52$). In this way, $E$ appears to be able to capture the insights of point symmetry and packing fraction in a single quantity.

\begin{table}[ht]
\small
  \caption{\ $12$-coordinate geometries listed by decreasing $E$. The $D_{3h}$ (order $12$) symmetry of the HCP coordination geometry (anticuboctahedral) is not to be confused with the $D_{6h}$ (order $24$) symmetry of the HCP lattice}
  \begin{tabular*}{0.48\textwidth}{@{\extracolsep{\fill}}llrrrr}
    \hline
    \multirow{2}{*}{Geometry} & \multicolumn{2}{c}{Point symmetry} & \multicolumn{3}{c}{Extracopularity} \\ \cmidrule{2-3} \cmidrule{4-6}
             & \multicolumn{1}{l}{$G$} & \multicolumn{1}{r}{$\abs{G}$} & \multicolumn{1}{c}{$k$} & \multicolumn{1}{c}{$m$} & \multicolumn{1}{c}{$E$} \\
    \hline
    ICO & $I_h$ & $120$ & 12 & 3 & 4.46 \\
    FCC & $O_h$ & $48$ & 12 & 4  & 4.04 \\
    HCP & $D_{3h}$ & $12$ & 12 & 6  & 3.46 \\
    BPP & $D_{5h}$ & $20$ & 12 & 7  & 3.24 \\
    \hline
  \end{tabular*}
  \label{table:12-coordinate-geometries}
\end{table}

\begin{table}[ht]
\small
  \caption{\ Cubic geometries listed by decreasing $E$}
  \begin{tabular*}{0.48\textwidth}{@{\extracolsep{\fill}}llrrrr}
    \hline
    \multirow{2}{*}{Geometry} & \multicolumn{2}{c}{Point symmetry} & \multicolumn{3}{c}{Extracopularity} \\ \cmidrule{2-3} \cmidrule{4-6}
             & \multicolumn{1}{l}{$G$} & \multicolumn{1}{r}{$\abs{G}$} & \multicolumn{1}{c}{$k$} & \multicolumn{1}{c}{$m$} & \multicolumn{1}{c}{$E$} \\
    \hline
    FCC & $O_h$ & $48$ & 12 & 4 & 4.04 \\
    BCC & $O_h$ & $48$ & 14 & 5 & 3.92 \\
    SC  & $O_h$ & $48$ & \phantom{1}6  & 2 & 2.91 \\
    \hline
  \end{tabular*}
  \label{table:cubic-geometries}
\end{table}

Table \ref{tab:cegs} expands the comparison to $22$ of the most commonly encountered coordination geometries in the physical sciences. Both $E$ and $\abs{G}$ are maximal for the icosahedral geometry but vary in their minima. For these $22$ geometries, $E$ is found to take $17$ distinct values while $\abs{G}$ is found to take $9$. With nearly double the resolution, $E$ appears to be the more capable classifier. Moreover, in those $5$ geometries with non-distinct $E$, differences in the bond angle distribution open up the possibility for an adjusted coefficient that can distinguish all $22$ geometries. By contrast, the presence of only $13$ distinct groups imposes a suboptimal hard limit on the resolution of symmetry-based approaches.

\section*{A problem inherent to symmetry}

\begin{figure}[!h]
    \centering
    \includegraphics[width=\linewidth]{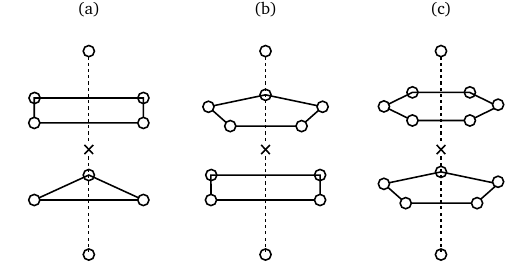}
    \caption{Three non-tessellating geometries with the same point group ($C_{s}$). Common molecules with this symmetry include water, formaldehyde, and sulphur dioxide.}
    \label{fig:nontessellating-geometries}
\end{figure}

Upon first glance, degeneracies of the kind exampled in Table \ref{table:cubic-geometries} might appear resolvable through the use of space groups, as these more comprehensive groups readily distinguish between the lattices implied by cubic coordination geometries. There are, however, two issues with this approach: Firstly, it assumes that all geometries with identical point symmetry can tessellate to form a lattice, which, as illustrated in \mbox{Fig. \ref{fig:nontessellating-geometries}}, is not the case. Secondly, and perhaps more fundamentally, the global structure of a crystal, as described by a space group, and its local structure around a single site, as described by a point group, fully determine each other and are thus informationally equivalent\cite{cover_2012} -- the presence of a difference in their symmetric descriptions is at odds with this equivalence, pointing to a problem inherent to our current, symmetry-based conception of structure at the particle scale.

\begin{table*}
    \centering
    \caption{\ Commonly encountered coordination geometries listed in order of increasing $E$}
    \begin{tabular}{lllrrrlr} \hline \multicolumn{1}{l}{Abbreviation} & \multicolumn{1}{l}{Coordination geometry} & \multicolumn{1}{l}{Polyhedral classification} & \multicolumn{1}{c}{$k$} & \multicolumn{1}{c}{$m$} & \multicolumn{1}{c}{$E$} & \multicolumn{1}{c}{$G$} & \multicolumn{1}{c}{$\abs{G}$} \\ \hline
    TBP & Trigonal bipyramidal & Deltahedral, bipyramidal & $5$ & $3$ & $1.737$ & $D_{3h}$ & $12$ \\
    SDS & Snub disphenoidal & Deltahedral & $8$ & $6$ & $2.222$ & $D_{2d}$ & $8$ \\
    CTP & Capped trigonal prismatic & Prismatic & $7$ & $4$ & $2.392$ & $C_{2v}$ & $4$ \\
    PBP & Pentagonal bipyramidal & Deltahedral, bipyramidal &  $7$ & $4$ & $2.392$ & $D_{5h}$ & $20$ \\
    BTP & Bicapped trigonal prismatic & Prismatic & $8$ & $5$ & $2.485$ & $C_{2v}$ & $4$ \\
    TET & Regular tetrahedral & Platonic, deltahedral & $4$  & $1$ & $2.585$ & $T_d$ & $24$ \\
    HBP & Hexagonal bipyramidal & Bipyramidal &  $8$ & $4$ & $2.807$ & $D_{6h}$ & $24$ \\ 
    CSA & Capped square antiprismatic & Antiprismatic & $9$ & $5$ & $2.848$ & $C_{4v}$ & $8$ \\
    CSP & Capped square prismatic & Prismatic & $9$ & $5$ & $2.848$ & $C_{4v}$ & $8$ \\
    TTP & Tricapped trigonal prismatic & Prismatic, deltahedral & $9$ & $5$ & $2.848$ & $D_{3h}$ & $12$ \\
    SC & Regular octahedral & Platonic, deltahedral, bipyramidal & $6$ & $2$ & $2.907$ & $O_h$ & $48$ \\
    BSA & Bicapped square antiprismatic & Deltahedral, antiprismatic & $10$ & $6$ & $2.907$ & $D_{4d}$ & $16$ \\
    BSP & Bicapped square prismatic & Prismatic & $10$ & $5$ & $3.170$ & $D_{4}$ & $8$ \\
    CPP & Capped pentagonal prismatic & Prismatic & $11$ & $6$ & $3.196$ & $C_{5v}$ & $10$ \\
    SA & Square antiprismatic & Antiprismatic & $8$ & $3$ & $3.222$ & $D_{4d}$ & $16$ \\
    HDR & Regular hexahedral & Platonic, prismatic & $8$ & $3$ & $3.222$ & $O_h$ & $48$ \\
    BPP & Bicapped pentagonal prismatic & Prismatic & $12$ & $7$ & $3.237$ & $D_{5}$ & $20$ \\
    HCP & Anticuboctahedral & Bicupolar & $12$ & $6$ & $3.459$ & $D_{3h}$ & $12$ \\
    BCC & Rhombic dodecahedral & Catalan & $14$ & $6$ & $3.923$ & $O_h$ & $48$  \\
    FCC & Cuboctahedral & Bicupolar & $12$ & $4$ & $4.044$ & $O_h$ & $48$ \\
    CPA & Capped pentagonal antiprismatic & Antiprismatic & $11$ & $3$ & $4.196$ & $C_{5v}$ & $10$ \\
    ICO & Regular icosahedral & Platonic, deltahedral, antiprismatic & $12$ & $3$ & $4.459$ & $I_h$ & $120$ \\ 
    \hline
    \end{tabular}
    \label{tab:cegs}
\end{table*}

\section*{Extracopularity generalises point symmetry}

All of what has so far been discussed seems to suggest that extracopularity generalises point symmetry -- that {every geometry with point symmetry has extracopularity, while not every geometry with extracopularity has point symmetry}. Under the idealisation of spherical neighbourhoods, this can in fact be proven.

\begin{figure}[!b]
    \centering
    \includegraphics[width=\linewidth]{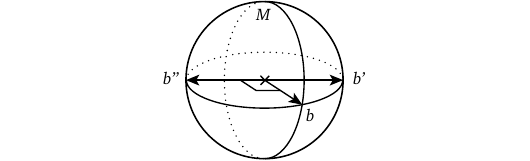}
    \caption{Introducing a fourth bond pointing anywhere off the meridional circle $M$ on this sphere trivialises point symmetry but not extracopularity.}
    \label{fig:converse-disproof}
\end{figure}

\begin{theorem*}
Consider a geometry of $k > 2$ bonds, equal in length. Then, nontrivial point symmetry ($\abs{G} > 1$) implies nontrivial extracopularity ($E > 0$), but not conversely.
\newpage
\begin{proof}
We first prove the implication, then disprove its converse.

\begin{proofpart} The implication is proven directly. 

Let $\abs{G} > 1$. Then, there exists a map $\mu \in G$ and two distinct bond pairs $\{b,b'\} \neq \{\beta, \beta'\}$ such that
\begin{equation}
    \begin{aligned}
    d(b, b') &= d\big[\mu(b), \mu(b')\big] \\
             &= d(\beta, \beta'),
    \end{aligned}
\end{equation}
where $d(b, b') := \norm{b-b'}$ is the Euclidean distance as in Fig. \ref{fig:trigonometry}.

Since all four bonds are equal in length, the two pairs also make the same angle,
\begin{equation}
    \begin{aligned}
        \angle(b, b') &= 2\arcsin\left[\frac{d(b, b')}{2\norm{b}}\right] \\
                      &= 2\arcsin\left[\frac{d(\beta, \beta')}{2\norm{\beta}}\right] \\ 
                      &= \angle(\beta, \beta').
    \end{aligned}
\end{equation}
But if two distinct bond pairs have the same angle, then the number of different bond angles is less than the total number of bond pairs, $m < n$. Recalling that $E = \log_2(n) - \log_2(m)$, we have
\begin{equation}
    E > 0.
\end{equation}
\end{proofpart}
\begin{proofpart}
The converse is disproven by counterexample.

Consider a geometry of $k>3$ bonds for which there exists a unique pair of distinct bond pairs $\{b,b'\} \neq \{b,b''\}$ such that
\begin{equation}
    d(b, b') = d(b, b'').
\end{equation}
This has the following two consequences:
\begin{itemize}
    \item There exists an angular equality. This implies that there are fewer bond angles than bond pairs, $m < n$. As a result,
    \begin{equation}
        E > 0.
    \end{equation}
    \item There does not exist a nontrivial isometry. Since point groups consist only of isometries, $G$ must be trivial. Thus,
    \begin{equation}
        \abs{G} = 1.
    \end{equation}
\end{itemize}
This counterexample is illustrated in Fig. \ref{fig:converse-disproof} 
\end{proofpart}
\end{proof}
\end{theorem*}

\section*{Discussion}

It is encouraging that a quantity based simply on the occurrence of fewer bond angles than combinatorially possible should lend itself to the characterisation of as complex a matter as structure. Through $E$, we observe the consistency of extracopularity with some of what is already known and understand that it generalises symmetry in a basic sense. This coefficient is, nevertheless, just one possible information-based quantifier of extracopularity, and indeed, structure.

There are two noteworthy challenges to the informational study of structure through $E$ in particular. Firstly, the number of different bond angles, $m$, can be difficult to ascertain in practice, with a variety of factors like thermal fluctuations ensuring that this is not as simple as counting the number of distinct angles.\cite{camkiran_2022b} Secondly, the extracopularity coefficient does not in its current form account for variation in bond lengths; such variation could constitute an important feature of structure, especially in higher entropy systems like liquids and polydisperse packings.

In addition to addressing the above limitations, it will be necessary to link structure as it is observed locally by $E$ to structure as it prevails globally to determine properties. Here, various statistical techniques present a promising avenue forward, in particular, simple descriptive statistics like the mean and variance, the joint distribution of neighbouring particle coefficients, and a coefficient correlation function similar in spirit to the radial distribution function.\cite{camkiran_2022b}

\section*{Concluding remarks}

Notwithstanding the challenges that yet lie ahead, it seems that a general, information-based theory of the particle-scale structure of materials is conceivable, if only in principle. This is in contrast to the prevailing, symmetry-based understanding, which for its many strengths, is fundamentally specialised to crystals; indeed, it is beset by crucial issues even in this comparatively simple setting. Ultimately, the study of structure through information constitutes a new bridge between the physical and statistical sciences. And if prior such links are to give any indication, it is that there is much to be gained by further work in this direction.

\section*{Conflicts of interest}
There are no conflicts to declare.

\section*{Acknowledgements}
The authors wish to acknowledge the contributions of Fabian Parsch to the development of the ideas discussed in this work.



\balance


\bibliography{main} 
\bibliographystyle{rsc} 

\end{document}